# Theoretical Analysis of Terahertz Detection of Resonant Tunneling Diodes

Masahiro Asada[1)] and Safumi Suzuki

Department of Electrical and Electronic Engineering, Tokyo Institute of Technology,

Meguro-ku, Tokyo, Japan.

1) E-mail: asada@pe.titech.ac.jp

Abstract:

We analyze the terahertz detection characteristics of resonant tunneling diodes (RTDs) using a set of simple equations that covers three detection modes; (i) direct detection, (ii) amplified detection, and (iii) self-homodyne (coherent) detection. (i) and (ii) are based on the square-law detection, and (iii) is on the homodyne detection with the RTD used as an injection-locked local oscillator. The calculated results exhibit small- and large-signal areas depending on irradiation power. In the small-signal area, the detection current is proportional to irradiated power for (i) and (ii), and to square root of irradiated power for (iii). The detection current has a peak at the bias voltage at the boundary between (ii) and (iii). Effect of frequency fluctuation of irradiated wave is analyzed for (iii), and it is shown that the detection current is proportional to irradiated power if the fluctuation becomes wider than injection-locking range. The analytical results in this paper reasonably explain the reported experiments.

The terahertz (THz) frequency band (roughly 0.1-10 THz) has a variety of potential applications, including imaging, spectroscopy, high-capacity communications, and radars [1][2]. Oscillators using resonant tunneling diodes (RTDs) are candidates for compact THz sources [3]-[6]. We have been studying RTD oscillators [6]. On the other hand, RTDs can detect THz waves [7]-[13]. It has been reported that the THz detectors using RTDs have three detection modes; direct detection, amplified detection, and self-homodyne (coherent) detection [12]. In this paper, we analyze the detection characteristics of RTDs using a set of simple equations that cover all three detection modes mentioned above.

The RTD detector in the present analysis is assumed to have the same structure as the RTD oscillator [6] in which an RTD is connected in parallel with a THz parallel resonance circuit.

Irradiation with a terahertz wave changes the dc current through RTD, resulting in the THz detection. The detection characteristics can be analyzed using an equivalent circuit shown in Fig. 1, regardless of whether RTD oscillates or not. The equivalent circuit in Fig. 1 is the same as that used in the analysis of the injection locking [15].

In Fig. 1, $L$ and $C$ are the inductance and capacitance of the resonance circuit dominated by the antenna and RTD, respectively, and $G_{ant}$ is the antenna conductance, which consists of the radiation conductance $G_{rad}$ and the loss conductance $G_{loss}$. $I_{in}$ and $\omega_{in}$ are the amplitude and angular frequency of the current source induced by the irradiated THz wave. $I_{in}$ is related to the irradiated power $P_{in}$ as $I_{in} = 2\sqrt{2\eta G_{rad} P_{in}}$ [14], where $\eta$ is the irradiation efficiency. $v(t) = V_{ac} e^{j(\omega_{in} t + \phi)}$ is the THz voltage across RTD with the amplitude $V_{ac}$ and the phase $\phi$ relative to that of the irradiated wave. The angular frequency of $v(t)$ is given by $\omega = \omega_{in} + d\phi/dt$.

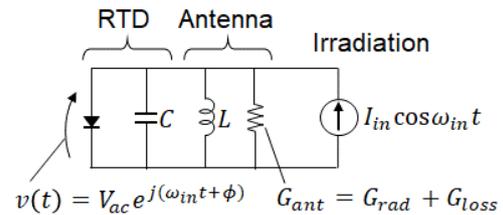

Fig.1: Equivalent circuit of RTD THz detector.

$V_{ac}$ and $\phi$ in $v(t)$ are obtained by the stationary solutions of the following equations [15].

$$\frac{dV_{ac}}{dt} - \alpha V_{ac} + \frac{\gamma}{4} V_{ac}^3 = I_{in} \cos\phi , \quad (1)$$

$$\frac{d\phi}{dt} + \omega_0 - \omega_{in} = -\frac{I_{in}}{2CV_{ac}} \sin\phi , \quad (2)$$

where $\alpha$ is the negative differential conductance (NDC) of RTD minus $G_{ant}$, i.e., $\alpha = -\partial I_{dc}/\partial V_{dc} - G_{ant}$ with $I_{dc}(V_{dc})$ being the I-V curve. Hereafter $I_{dc}(V_{dc})$ is approximated by a cubic function as $I_{dc}(V_{dc}) = -a(V_{dc} - V_c) + b(V_{dc} - V_c)^3 + I_c$ [16], where $a$ and $b$ are constants, and $V_c$ and $I_c = -aV_c + bV_c^3$ are the voltage and current at the center of the NDC region. By this approximation, $\alpha$ is given by $\alpha = a - G_{ant} - 3b(V_{dc} - V_c)^2$. $\gamma$ in eq. (1) is equal to $3b$, and $\omega_0$ in eq. (2) is the angular resonance frequency of the resonance circuit in Fig. 1 given by $1/\sqrt{LC}$, which is almost equal to the oscillation angular frequency if RTD oscillates. Although the precise analysis shows that the oscillation frequency slightly deviates from the resonance frequency [18], we neglect this deviation.

Detection modes of RTD can be classified into the following three types according to the value of $\alpha$ depending on the bias voltage $V_{dc}$.

(i) Direct detection: $\alpha + G_{ant} < 0$. In this mode, $V_{dc}$ is set outside the NDC region.

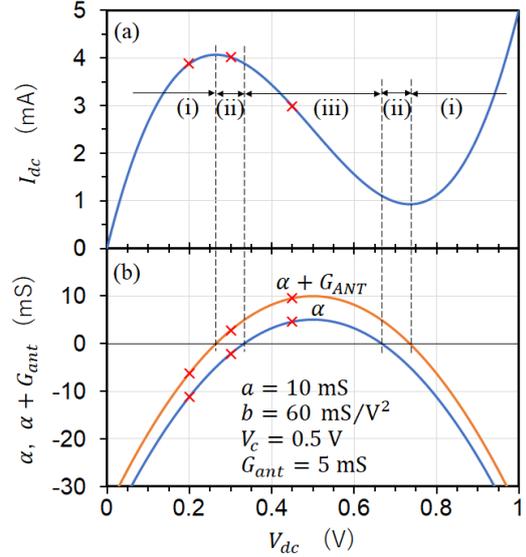

Fig.2: Bias voltage regions for detection mode (i)-(iii) in (a) I-V curve approximated by cubic function and (b) $\alpha$ and $\alpha + G_{ant}$. Red x points indicate bias voltages at which the detection characteristics are calculated below in Fig. 3.

(ii) Amplified detection: $0 \leq \alpha + G_{ant}$ and $\alpha < 0$. In this mode, $V_{dc}$ is set inside the NDC region, but RTD is not oscillating.

(iii) Self-homodyne (coherent) detection: $0 < \alpha$. RTD is oscillating in this mode.

Figure 2 shows the bias voltage regions for the three detection modes (i)-(iii) in the curves of $I_{dc}(V_{dc})$, $\alpha$, and $\alpha + G_{ant}$. Red x points indicate the points at which the detection characteristics are calculated in Fig. 3 below. These are located at $V_{dc}$ =0.2, 0.3, and 0.45 V in the regions (i), (ii), and (iii), respectively.

In the calculation below, we assume $\omega_{in} = \omega_0$ in eq. (2). This results in $\phi$=0 in the stationary state. However, even if the difference $\Delta\omega = |\omega_0 - \omega_{in}|$ is nonzero, $\phi$ can be well approximated by zero ($\phi \lesssim 10^{-2}$ for $\Delta\omega/\omega_0 \lesssim 10\%$ and the parameter values used in the calculation).

In the detection mode (iii), $v(t)$ exists without the irradiation because RTD is oscillating, and changes with the irradiation. We assume that the free-running oscillation frequency of RTD is set nearly equal to the irradiation frequency, i.e., $\omega = \omega_0 \simeq \omega_{in}$. RTD is thus injection-locked to the irradiation [10][11]. In this condition, $\phi$ is almost equal to zero from eq. (2), and the injection locking of RTD is maintained even at very low irradiation power. However, if the linewidth of the irradiation wave or oscillation wave of RTD is finite, there is a lower limit to the irradiation power for the injection locking, as discussed later.

Due to the generation or change of $V_{ac}$ in eq. (1) by the irradiation, the dc current through RTD changes, and the irradiation is detected. The dc current change $\Delta I_{det}$ is calculated as follows [16].

For the detection modes (i) and (ii), $\Delta I_{det}$ is calculated as

$$\Delta I_{det} = \frac{3b}{2}(V_{dc} - V_c)V_{ac}^2. \qquad (3)$$

This equals to the square-law detection ( = $(\partial^2 I_{dc}/\partial V_{dc}^2)V_{ac}^2/4$). In the detection mode (ii), $V_{ac}$ is amplified by the NDC of RTD. The square-law detection is the limit of low photon energy of the photon-assisted tunneling [8].

For the detection mode (iii), $\Delta I_{det}$ is calculated as

$$\Delta I_{det} = \frac{3b}{2}(V_{dc} - V_c)\left(V_{ac}^2 - \frac{4\alpha}{\gamma}\right). \qquad (4)$$

In eq. (4), the second term $4\alpha/\gamma$ of the second brackets is equal to $V_{ac}^2$ due to oscillation only, and it is subtracted from the first term which is the total $V_{ac}^2$ generated by both oscillation and irradiation. The second brackets in eq. (4) create a cross term between $V_{ac}$ due to the oscillation ($\sqrt{4\alpha/\gamma}$) and $V_{ac}$ due to the irradiation, and thus, the homodyne detection is obtained with the oscillation of RTD being the local oscillator.

Figure 3 shows the absolute values of detection current $\Delta I_{det}$ calculated with eqs. (1)-(4) as a function of irradiated power $P_{in}$. The bias voltages for the detection modes (i)-(iii) are indicated in Fig.2 by x points. The dashed line is the case in (iii) where the irradiated wave is assumed to fluctuate with the standard deviation of 100 MHz, as discussed later.

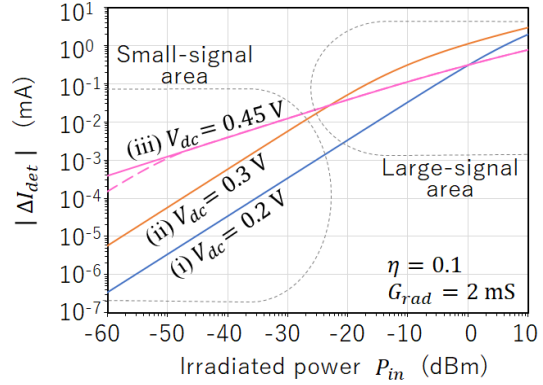

Fig. 3: Calculated detection current as a function of irradiated power. Bias voltages for detection modes (i)-(iii) are indicated in Fig. 2 as x points. The dashed line is the case in (iii) where the irradiated frequency is assumed to fluctuate with the standard deviation of 100 MHz, as discussed later.

The $P_{in}$ dependence of $|\Delta I_{det}|$ can be roughly divided into small-signal and large-signal arears depending on $P_{in}$. The boundary of $P_{in}$ between these two arears is approximately -20 – -30 dBm in Fig. 3, although which depends on the irradiation efficiency $\eta$ and radiation conductance $G_{rad}$.

In the small-signal area in Fig. 3, $|\Delta I_{det}| \propto P_{in}$ for the detection modes (i) and (ii), and $|\Delta I_{det}| \propto P_{in}^{1/2}$ for (iii). These results have been obtained in the experiments [10][11].

In the large-signal area, $|\Delta I_{det}|$ gradually saturates with increasing $P_{in}$. The detection mode (ii) shows the highest saturation tendency. In the limit of large $P_{in}$, $|\Delta I_{det}| \propto P_{in}^{1/3}$ is resulted from eqs. (1)-(4) for all detection modes. The saturation of $|\Delta I_{det}|$ arises from the nonlinear dependence of $V_{ac}$ on $I_{in}$ in eq. (1). The large-signal area corresponds to the experiment in [13], and qualitative agreement was obtained. (Since the detected power was used instead of the detection current in [13], the slope of the measured curves is twice the slope of the detection current.) The measurement result in [13] for the detection mode (ii) seems to include the transient region from the small-signal area to the large-signal area. The $P_{in}$ dependence of the detected power in this region is larger than expected from Fig. 3. The reason for this discrepancy is currently unknown. A possible reason could be, for example, that the actual *I-V* curve of RTD cannot be well approximated by the cubic curve.

Figure 4 shows $|\Delta I_{det}|$ as a function of $V_{dc}$. $|\Delta I_{det}|$ has a peak at the boundary between the detection modes (ii) and (iii), where $\alpha = 0$, and becomes zero at the center of the NDC region. The sensitivity in the detection mode (i) is always smaller than that in (ii).

The irradiated wave has a finite spectral linewidth due to the noise in the irradiation source or the modulation by an external signal. In the detection mode (iii), the injection-locking range may become narrower than the spectral linewidth of the irradiated wave as $P_{in}$ decreases. Here we discuss the detection current of the irradiated wave with a finite linewidth. As a simple case, we assume that the amplitude of the irradiated wave is constant, but the frequency varies due to modulation by an external signal, resulting in a finite linewidth.

In this case, the frequency of the irradiation wave moves in and out of the locking range if $P_{in}$ is low. Although the frequency movement depends on external signal, we assume for simplicity that the irradiation frequency varies with the following Gaussian distribution which represents the probability such that the irradiated wave has the frequency $f_{in} + \delta f_{in}$ around the average $f_{in}$.

$$p(\delta f_{in}) = \frac{1}{\sqrt{2\pi}\Delta f_{in}} \exp\left(-\frac{\delta f_{in}^2}{2\Delta f_{in}^2}\right), \quad (5)$$

where $\Delta f_{in}$ is the standard deviation of the frequency variation which is proportional to the linewidth of the irradiated wave.

The probability that the irradiation frequency is within the locking range is obtained as

$$P = \int_{-\Delta f_{lock}}^{\Delta f_{lock}} p(\delta f_{in})\, d(\delta f_{in}) = \mathrm{erf}\left(\frac{\Delta f_{lock}}{\sqrt{2}\Delta f_{in}}\right), \quad (6)$$

where $\Delta f_{lock}$ is the locking range approximately given by [15] $(f_{in}/Q_{rad})\sqrt{P_{in}/P_{out}}$. $Q_{rad} \simeq 2\pi f_{in} C/G_{rad}$ is the Q factor of radiation, $P_{out} = (G_{rad}/2)(4\alpha/\gamma)$ is the output power of RTD, and erf is the error function.

When the irradiation frequency goes out of the locking range, the component of the irradiation frequency included in the oscillation wave of RTD will rapidly decreases, and it becomes impossible to observe the change in dc current by the homodyne detection. Thus, the detection current under the variation of the irradiation frequency is obtained by the product between eqs. (4) and (6). An example of this detection current is indicated by the dashed line in Fig. 3 for $\Delta f_{in} = 100$ MHz, $C = 20$ fF, and $G_{rad} = 2$ mS (the factor $\Delta f_{in} C/G_{rad}$ discussed later is $10^{-3}$).

If $P_{in}$ is large enough that the locking range is much wider than the fluctuation ($\Delta f_{lock} \gg \Delta f_{in}$), equation (6) is nearly equal to 1. In this case, the detection current is not influenced by the frequency variation, and is given almost solely by eq. (4) ($\propto P_{in}^{1/2}$). If $P_{in}$ is so small that the locking range is much narrower than the fluctuation ($\Delta f_{lock} \ll \Delta f_{in}$), equation (6) is approximately proportional to $P_{in}^{1/2}$. Equation (4) is also proportional to $P_{in}^{1/2}$, and thus, the detection current (eq. (4) × eq. (6)) is proportional to $P_{in}$.

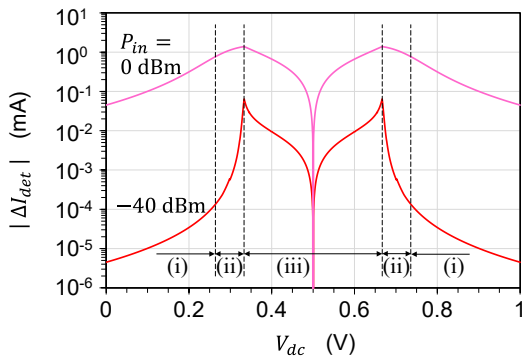

Fig. 4: Calculated detection current as a function of bias voltage.

$P_{in}$ at the boundary between $|\Delta I_{det}| \propto P_{in}^{1/2}$ and $|\Delta I_{det}| \propto P_{in}$ is estimated from the condition $\Delta f_{lock} \simeq \Delta f_{in}$, which is proportional to $(\Delta f_{in} C/G_{rad})^2$. Figure 5 shows the absolute value of the detection current for different values of $\Delta f_{in} C/G_{rad}$. As shown in Fig. 5, the boundary of $P_{in}$ discussed above largely changes with $\Delta f_{in} C/G_{rad}$, and a small value of this factor is required to maintain high sensitivity in a wide range of $P_{in}$ for the detection mode (iii).

If the modulation of the irradiated wave induces simultaneous presence of many spectral lines in addition to the frequency variation, a different analysis will be required.

If the irradiated wave is not modulated by an external signal, the linewidth of the irradiated wave is usually caused by the phase noise. The influence of linewidth in this case can be considered as follows. When a phase jump occurs in the irradiated wave due to noise, the phase of the oscillation wave of RTD which is injection-locked to the irradiation wave follows to the phase of the irradiation wave within the time constant $\tau_{lock} \sim 1/\Delta f_{lock}$, as seen from eq. (2). On the other hand, the average time between two phase jumps is $\tau_{coh} \sim 1/\Delta f$.

If $\tau_{lock} \ll \tau_{coh}$, i.e., $\Delta f \ll \Delta f_{lock}$, the phase of the oscillation wave of RTD immediately catches up with the phase jump of the irradiated wave. This means that the phase difference between the irradiation wave and the oscillation wave ($\phi$ in eq. (2)) immediately vanishes after every phase jump of the irradiated wave, because the oscillation frequency is equal to the irradiation frequency. As a result, there is almost no effect of phase jump.

If $\tau_{lock} > \tau_{coh}$, i.e., $\Delta f > \Delta f_{lock}$, before the phase of the oscillation wave of RTD catches up with the first phase jump, the next phase jump occurs. In this case, a nonzero value of $\phi$ always exists. Since $V_{ac}$ in eq. (1) responds sufficiently fast to the variation of $\phi$, the average of $V_{ac}$ is smaller than $V_{ac}$ for $\phi = 0$ through the factor $\cos\phi$ (<1) in eq. (1). Consequently, the detection current is smaller than that in the case of $\Delta f \ll \Delta f_{lock}$. The dependence of $|\Delta I_{det}|$ on $P_{in}$ will change at the irradiation power approximately satisfying $\Delta f \sim \Delta f_{lock}$, as in the above case where the irradiated wave is modulated.

The above qualitative discussion should be able to be quantitatively analyzed by adding the phase jump (random noise) of the irradiated wave to eqs. (1) and (2) and finding the average value of the changes in $V_{ac}$. However, mathematical treatment is complicated and remains as a future subject. In the experiment [11], the dependence of the detection current on the irradiated power changes around the irradiation power satisfying $\Delta f \sim \Delta f_{lock}$.

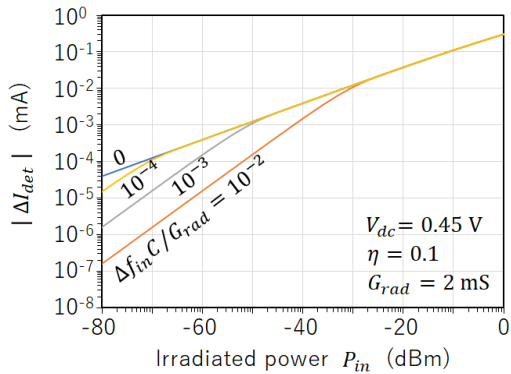

Fig. 5: Calculated detection current as a function of irradiated power for detection mode (iii) at bias voltage of 0.45 V with different values of $\Delta f_{in} C/G_{rad}$.

A separate analysis will also be required if the linewidth of RTD is wider than the locking range. In this case, it is necessary to add the noise term of RTD [17] into the equation for injection locking.

In conclusion, we analyzed the detection characteristics of RTDs using a set of simple equations that cover three detection modes; (i) direct detection, (ii) amplified detection, and (iii) self-homodyne (coherent) detection. The calculated results exhibited small- and large-signal areas depending on irradiation power. In the small-signal area, the detection current is proportional to irradiated power for the modes (i) and (ii), and to square root of irradiated power for the mode (iii). The detection current has a peak at the bias voltage at the boundary between (ii) and (iii). Effect of frequency variation of irradiated wave was also analyzed for the mode (iii), and it was shown that the detection current is proportional to irradiated power if the frequency varies wider than injection-locking range. The analytical results in this paper reasonably explain the reported experiments.

Acknowledgement
The authors thank Honorary Prof. Y. Suematsu, Emeritus Profs. K. Furuya and S. Arai, Profs. Y. Miyamoto and N. Nishiyama, and Assoc. Prof. M. Watanabe of the Tokyo Institute of Technology for their continuous encouragement. This study was supported by a scientific grant-in-aid (21H04552) from JSPS, CREST (JPMJCR21C4) from JST, X-NICS (JPJ011438) from MEXT, SCOPE (JP215003005) from MIC, commissioned research from NICT (No. 03001), and the Canon Foundation.